# First principles design of 2 dimensional Nickel dichalcogenide Janus materials NiXY (X,Y=S,Se,Te; X≠Y)

A. Sengupta[1*]

[1] University of North Bengal, P.O. NBU, Dist. Darjeeling, WB- 734 013, India

*Corresponding author e-mail: amretashis@nbu.ac.in

*Abstract:* In this work, we propose novel two-dimensional (2D) Janus Ni dichalcogenide materials and explore their feasibility, stability and evaluate their electronic and optical properties with ab-initio calculations. Three unique Janus materials, namely NiSSe, NiSTe and NiSeTe, with 2H hexagonal structure were proposed. Density functional theory (DFT) calculations, show that among the three proposed NiXY Janus 2D materials, NiSSe had the best energetic and dynamical stability. GGA PBE calculations showed NiSSe to have a semi-metallic bandstructure with the Ni-Se interaction having a dominant role in the band profile near the Fermi energy. Electron localization function (ELF) and total potential plots show a distinguishable asymmetry in terms of valence electron localization and distribution between the S and Se atoms in 2D NiSSe. The presence of large amount of electron gas like feature in the ELF around the chalcogen atoms also indicates their importance in the conduction properties. Optical properties calculated with random phase approximation (RPA) show the 2D NiSSe to have broad spectrum optical response with significant peaks lying in each of the infra-red, visible and the ultraviolet range of the spectra.

## I. Introduction

In recent years there has been a good amount of interest generated in two dimensional (2D) structures with chemically asymmetric bottom and top surfaces, known as Janus 2D materials. [1]-[3] The varied surface chemistry offers interesting properties related to piezoelectricity, catalysis, water splitting, Rashba spin splitting to name only a few. [1]-[7] Originally starting out with graphene sheets differently functionalized on either surfaces [1,2], a vast number of 2D Janus materials, such as MoSSe , WSS, MoSTe, WSeTe , VSSe , PdSSe , HTaSe2F , InGaSSe ,SnSSe and many more have been reported so-far. [1]-[7] With the ever growing research and development activity in the 2D material scenario, the computational design efforts of such novel Janus 2D materials is worth pursuing to compliment future experimental efforts and also act as a pointer for the same.

Of late, Ni based 2D materials such as $NiCl_2$, $NiBr_2$, $NiI_2$, $NiS_2$ , $NiSe_2$, $NiTe_2$ [8]-[11] etc. have been predicted as suitable candidates for various applications in batteries, supercapacitors, electronic/spintronic devices, electrocatalysis etc. [8]-[13] Among these materials, the layered Nickel dichalcogenides [7]-[10], have the 2H hexagonal structure similar to $MoS_2$ and many other transition metal dichalcogenides (TMDC). This presents us with an opportunity to propose and design of a number of Janus 2D material with the general formula NiXY (X,Y=S,Se,Te; X≠Y).

In the present work we propose and investigate from first principles calculations the properties of a number of such Janus 2D Ni dichalcogenide materials. We propose Janus NiSSe, NiSeTe, and NiSTe monolayers, having 2H hexagonal structures with different chalcogens on the top and the bottom surfaces. With density functional theory calculations, we evaluate the stability, structural properties and electronic and optical properties of the materials under consideration.

## II. Methods

The density functional theory (DFT) calculations are carried out with the Quantum ESPRESSO [14, 15] code as implemented in Material Square. [16] We employ generalized gradient approximation (GGA) DFT with Perdew-Burke-Ernzerhoff (PBE) exchange and correlation functional for our calculations. [17] The 2D hexagonal unit cells with vacuum padding of 15Å on each side were sampled with 8x8x1 Monkhorst-pack k-point grid. [18]. The cut-off energies for wave function was taken as 30Ry and that for the density was taken as 180Ry. Solid state pseudopotentials library (SSSP) precision pseudopotentials with scalar relativistic corrections were used for the calculations. [19] Davidson diagonalization algorithm [20] was used for the self-consistent calculations with electron convergence threshold of $10^{-6}$ Ry. The Broyden-Fletcher-Goldfarb-Shanno (BFGS) algorithm [21] was utilized for optimization calculation to minimize the Hellman-Feynmann forces below 0.01eV/Å.

The formation energy / cohesive energy for the structures was defined as follows

$$E_{coh} = -(E_{NiXY} - E_{Ni} - E_X - E_Y) \tag{1.1}$$

In the above equation, $E_{NiXY}$ indicates total energy of the proposed Janus material, and $E_{Ni,X,Y}$ indicates the total energy for the constituent atoms. Considering the sign convention used a net positive value of the cohesive energy indicates a favourable outcome regarding stability.

In order to study the dynamical stability of the proposed Janus 2D materials, we calculate the phonon bandstructure and phonon density of states for the optimized structures, with density functional perturbation theory (DFPT) calculations in ESPRESSO. [22] The optical properties of the proposed Janus 2D material, was obtained with random phase approximation (RPA) calculations with the epsilon.x package of the ESPRESSO suite as implemented in Materials Square [23,24].VESTA software [25] is used for the visualization purposes of the ELF and potential data. All the other plots and structure visualizations are carried out with the Materials Square integrated GUI. [16]

## III. Results & Discussions

The structure of the proposed Janus 2D materials are shown in Fig. 1, with NiSSe as the representative example. Here it should be mentioned that in Fig. 1, we have shown the supercell representation of the optimized structure for visualization purposes, with the hexagonal unit cell used for the calculations being highlighted. We consider a 2H hexagonal structure of $NiX_2$ as with one plane of the chalcogen X being replaced by Y to create a Janus material. Considering S, Se and Te chalcogen atoms, we can obtain three unique Janus Ni TMDC structures, namely NiSSe, NiSTe and NiSeTe. As we consider only the monolayer form of the Janus materials, the choice of order of the top/ bottom layer of chalcogenides (i.e. NiXY or NiYX) is deemed equivalent.

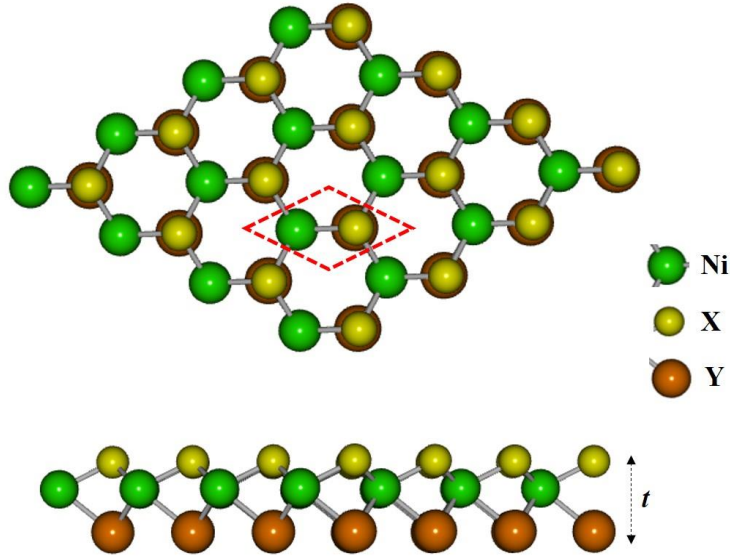

**Fig. 1:** Structure of the proposed NiXY (X,Y=S,Se,Te; X≠Y) Janus 2D material, the dashed red line showing the hexagonal unit cell of the material (used for the calculations). The top view (001) and the side view (110) of the structure is presented herein. For convenience a repeated supercell is shown.

The calculations show that among the Janus materials studied, NiSSe had the highest cohesive energy of 1.68eV followed by NiSTe (1.48eV) and NiSeTe (1.13eV), which indicates favourable condition for formation of these materials. The optimized lattice parameters, the bond lengths between Ni and the first (X) and second (Y) chalcogen, according to formula, the thickness t (i.e. the distance between the bottom and top chalcogen) and the calculated cohesive energies are listed in Table I. The structural parameters for the pure Ni dichalcogenides $NiS_2$, $NiSe_2$ and $NiTe_2$ are provided for reference in Table II.

**Table I.** The calculated structural parameters of the optimized NiXY Janus 2D material

| Formula | a (Å) | b (Å) | $d_{Ni-X}$ (Å) | $d_{Ni-Y}$ (Å) | t (Å) | $E_{coh}$ (eV) |
|---|---|---|---|---|---|---|
| NiSSe | 3.542 | 3.542 | 2.316 | 2.497 | 2.520 | 1.68 |
| NiSTe | 3.543 | 3.543 | 2.294 | 2.630 | 2.693 | 1.48 |
| NiSeTe | 3.541 | 3.541 | 2.433 | 2.569 | 2.871 | 1.13 |

**Table II.** The calculated structural parameters of the optimized 2D $NiX_2$

| Formula | a (Å) | b (Å) | d (Å) | t (Å) |
|---|---|---|---|---|
| NiS2 | 3.542 | 3.542 | 2.307 | 2.135 |
| NiSe2 | 3.507 | 3.507 | 2.417 | 2.639 |
| NiTe2 | 4.166 | 4.166 | 2.416 | 2.637 |

In order to evaluate the dynamical stability of the proposed Janus 2D materials, we calculate the phonon bandstructure and phonon density of states for the optimized structures, with density functional perturbation theory (DFPT) calculations in ESPRESSO. [22] Our calculations show that NiSSe has a very good dynamical stability in terms of the absence of any negative phonon bands as indicated by the phonon bandstructures and phonon DOS shown in Fig. 2.

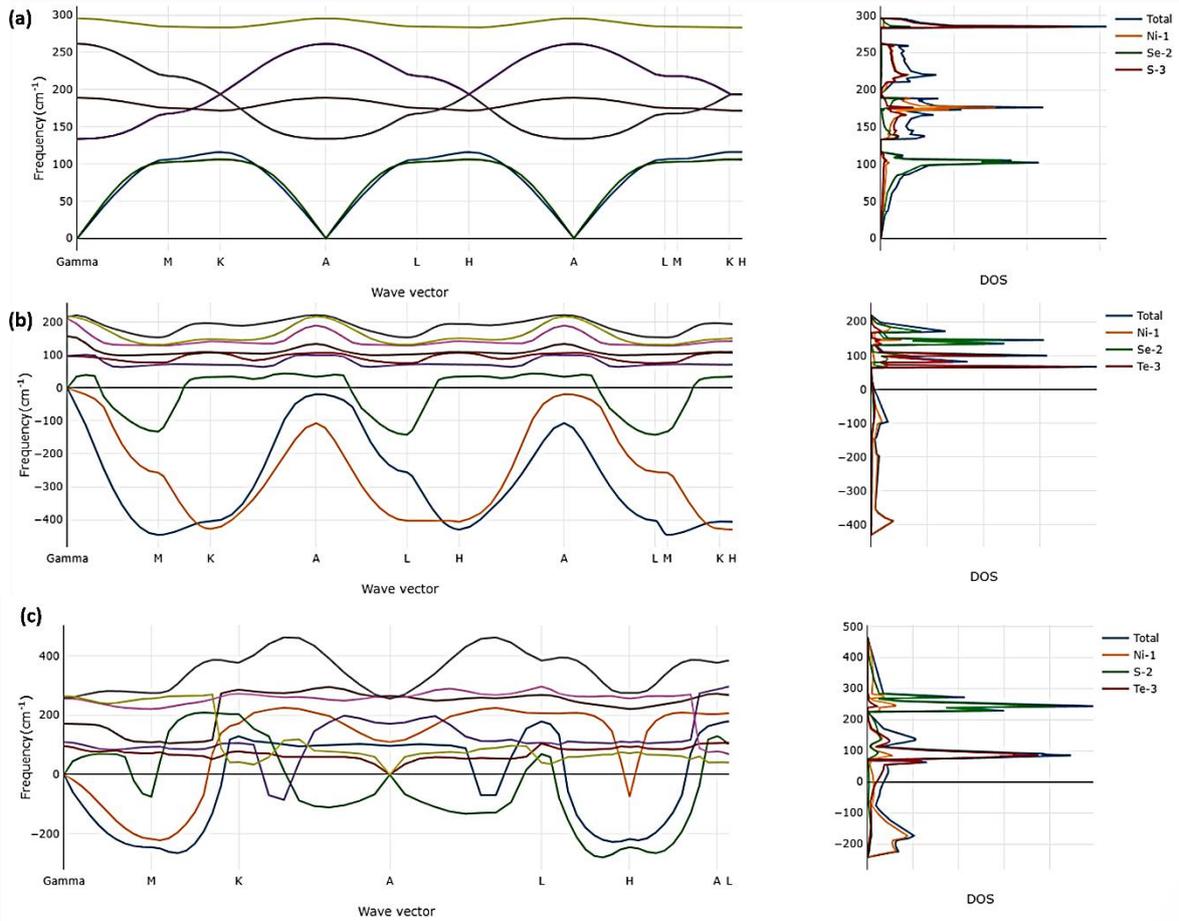

**Fig. 2:** The calculated phonon dispersion and the phonon density of states (DOS) plots for the proposed Janus 2D NiXY materials- **(a)** NiSSe **(b)** NiSeTe and **(c)** NiSTe

However the same is not true for NiSeTe and NiSTe as the presence of negative energy states as evident from the phonon dispersion and the phonon DOS, indicate for dynamical instability in the structures. This being the case, we limit our further discussions mostly to the Janus NiSSe, which shows stable nature from the calculations in its free standing form. Some results for NiSTe and NiSeTe are still provided in the supplementary material [26] for the reader's topical interest.

The calculated electronic bandstructure and electronic density of states of 2D NiSSe is presented in Fig. 3. It is observed in Fig. 3(a) that NiSSe has a semi-metallic nature by virtue of a number of bands crossing the Fermi energy. The zero-gap nature of the Janus material is also seen in 2D $NiS_2$ which shows a zero energy band gap, though with a different band profile. [27] A comparison of the bandstrcture of the proposed 2D NiSSe with that of 2H phase 2D $NiS_2$ and $NiSe_2$ is shown in Fig. 3(b). From Fig. 3(b) it can be seen that the Janus material bands are more influenced by the Se atoms, especially near the Γ point as compared to that by the S atoms, as they more closely follow the band profile of $NiSe_2$ than $NiS_2$ in these regions.

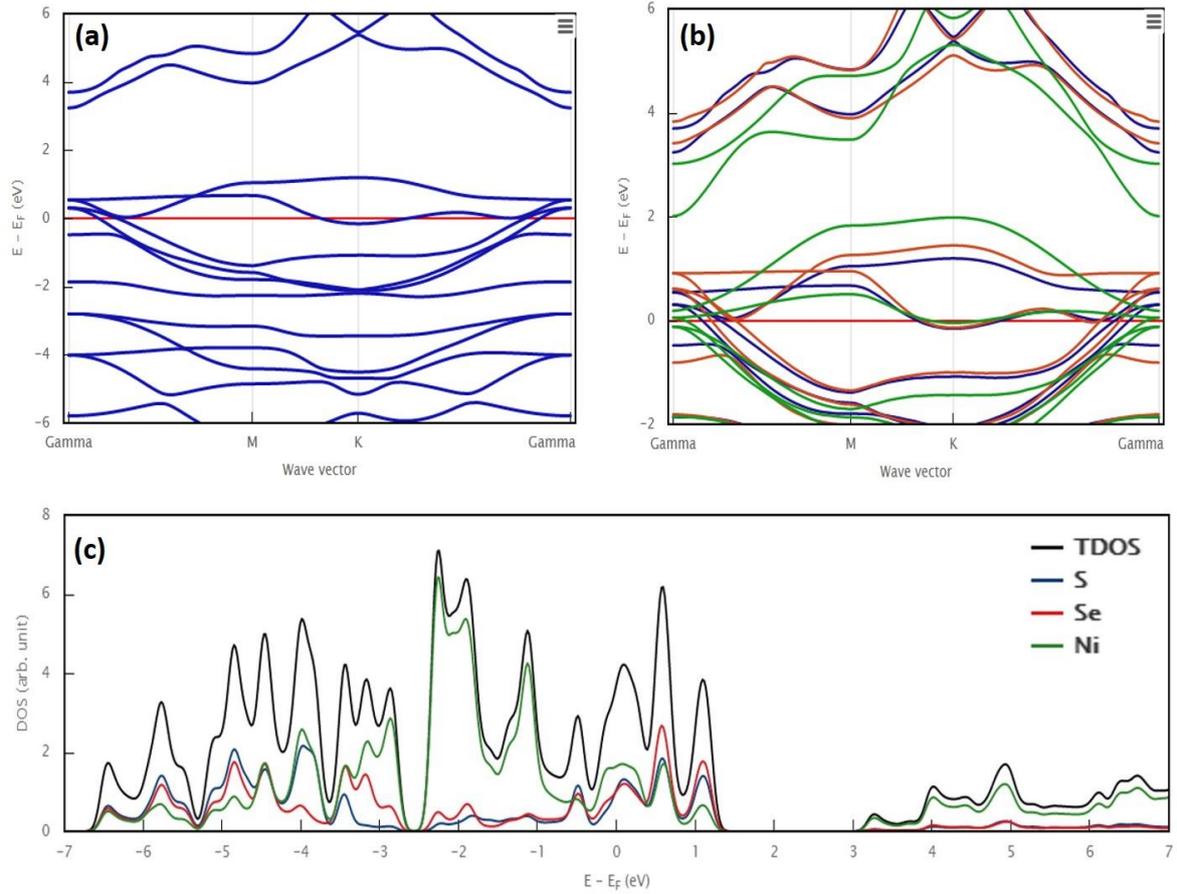

**Fig. 3: (a)** Calculated bandstructure of NiSSe **(b)** comparison of bandstructure of 2D Janus NiSSe (blue line) with 2D NiSe$_2$ (orange line) and 2D NiS$_2$ (green line) and **(c)** density of states of the 2D Janus NiSSe

The second band above the Fermi energy in between the M and K point is pushed down as compared to NiS$_2$ owing to the greater contribution from the Se 4s and 4p orbitals as compared to the S 3s and 3p orbitals. The bands at higher energy, in particular the ones near 3-4eV, have most contribution from the Ni atom 3s and 4s orbitals. Hence, it can be understood that the shape of these bands are affected not directly from the Se states, but rather the Ni-Se interactions, as they are much close to those due to NiSe$_2$. The contributions by different elements and states are more elaborately shown in the projected bands in the supplementary information. [26] The total density of states (TDOS) along with the contribution of the constituent elements in the density of states is shown in Fig. 3(c). The larger contribution of Se as compared to S near the Fermi level, and that of Ni at higher energies is also corroborated from the DOS plot as well.

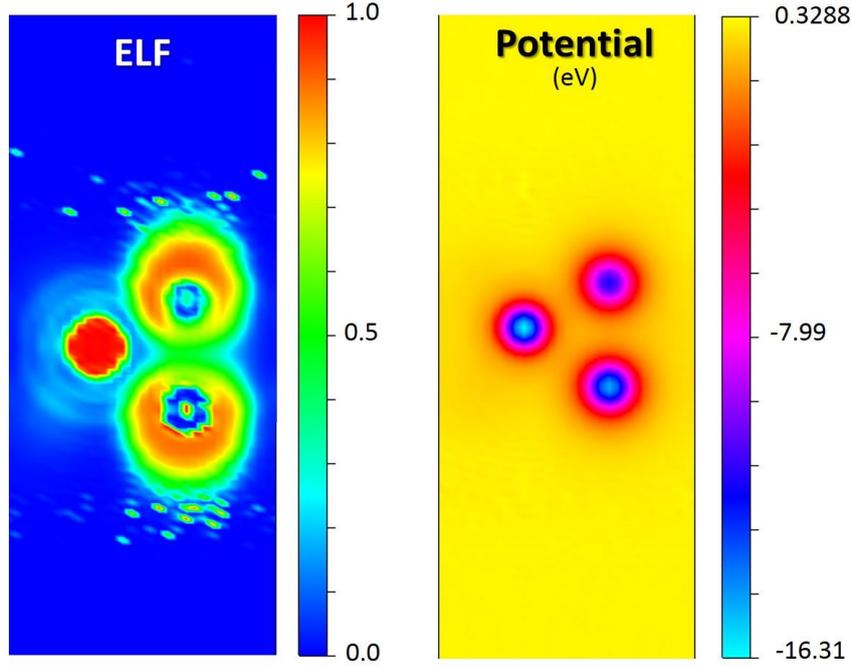

**Fig. 4:** The electron localization function (ELF) and the total potential plots of 2D NiSSe unit-cell shown in the (110) cut plane.

An important aspect of the Janus 2D structure can be the spatial distribution of the valence electrons which can be understood from the electron localization function (ELF). In Fig. 4, we present the calculated ELF and the total potential (inclusive of the exchange-correlation energy) of the 2D NiSSe. The plots are shown in the (110) plane of the hexagonal unit cell of NiSSe, the top chalcogen atom in the 2H-NiSSe is S and the bottom one being Se, as also shown in Fig. 1. To briefly discuss the background of the interpretation of the ELF, a value of 1.0 (i.e. red in the BGR colorbar) indicates electrons most likely to be localized in that region, while a value of 0.0 (i.e. blue) in a region signifies electrons least likely to be localized there. [28] The possibility of having a feature resembling an electron gas, important for transport and also screening effect, can be seen from the green coloured regions (which have a value of about 0.5). [28] From the plot in Fig. 4, it is seen that a strong localization of electrons occurs at the site of the metal atom, while moderate localization regions (yellow-red) appear around the chalcogen atoms. The electrons around the chalcogen atoms tend to spread away from the each other forming an elongated elliptical shape surrounding the chalcogen atom cores. The electrons tend to have a higher localization around the Se atom seen from the more red spots around it, than the S atom. Surrounding these regions and also in the space between the two chalcogen atoms there exists a clear presence of electron gas, which may promote carrier transport through these regions. As seen from the potential plot the Ni and Se atoms have a more negative potential as compared to the S atom, as indicated by the blue-cyan coloured spots in the CMY colormap used. In an overall positive environment the atoms act as negative potential centres, and the asymmetry in potential between the top and the bottom chalcogen layers (i.e. the S and Se atoms), indicates a possible change in the electric dipole moment in the Janus material as compared to pure $NiS_2$ and $NiSe_2$, where no such asymmetry exists. This asymmetry is consistent with the findings of the ELF plot and can be of importance considering applications such as catalysis and water splitting with 2D materials. The ELF and polarization plots for pure $NiS_2$ and $NiSe_2$ are also included in the supplementary information for comparison. [26]

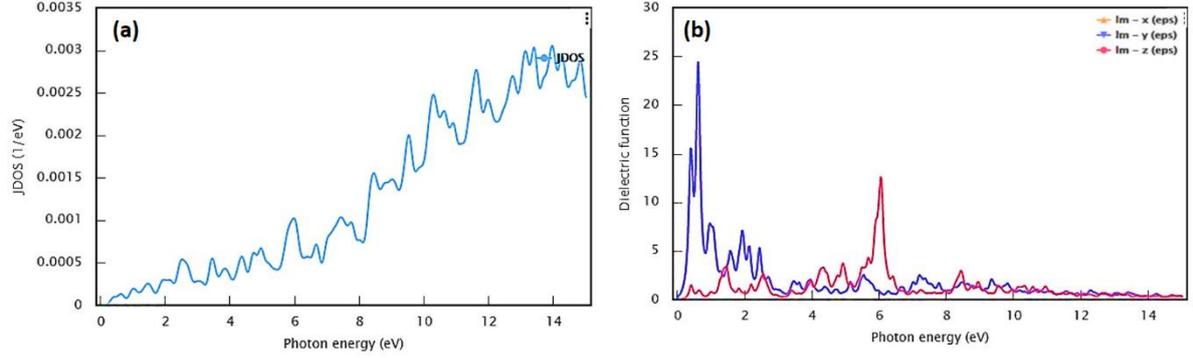

**Fig. 5: (a)** The joint density of states (JDOS) and **(b)** the imaginary dielectric function ($\varepsilon_2$) of 2D NiSSe obtained from our DFT calculations.

The optical properties of the proposed Janus 2D material, was obtained with random phase approximation (RPA) calculations with the epsilon.x package [23] of the ESPRESSO suite as implemented in Materials Square [16]. We present here in Fig. 5(a) the joint density of states (JDOS) plot of the material which is an indicator of availability of states for optical transitions to take place for a given photon energy and in Fig. 5(b) the imaginary part of the dielectric function which is an excellent indicator of the absorption properties of the material. The JDOS plot indicates that NiSSe has possible absorption states even for low energy photons in the infrared range and a significant peak around 2.7eV. More peaks are observed in the UV and beyond ranges. As for the dielectric function, we have presented all three (x, y and z) components of $\varepsilon_2$. As a monolayer 2D material the optical properties of NiSSe would have a strong dependence on the imaginary z component of the dielectric function, which is indicated by the red line in the dielectric function plot. This component, in the UV-Vis-NIR range has peaks around 0.65, 1.54, 2.45 and 2.73eV, which lie in the IR, NIR, visible (Green), and ultraviolet (UV) frequencies respectively. Hence a broad spectrum response can also be expected from the proposed Janus 2D material.

### IV. Conclusion

In this work, we carry out ab-initio studies on a proposed novel two-dimensional (2D) Janus Ni dichalcogenide material. With PBE calculations in DFT we and explore the energetic and synamical stability, and found NiSSe to be the most stable of the proposed Ni based Janus TMDC. Results showed NiSSe to have a semi-metallic band structure influenced more by the Ni-Se interaction as compared to Ni-S. ELF and total potential plots reveal a dissimilar amount of carrier distribution among the chalcogen atoms, with a strong presence of electron gas, which is significant from quantum transport perspectives. Also wide spectrum optical response was predicted for NiSSe from the RPA calculation results for imaginary dielectric function.
Overall, the calculation results presented herein indicates a good prospect for a proposed Ni based Janus 2D chalcogenide (i.e. NiSSe) from the standpoint of novelty, stability and properties. This being said it is our expectation that more research and development especially with the experimental aspect would be performed by other groups on synthesizing NiSSe and other such materials in the near future.

## Acknowledgement

The author thanks the Science and Engineering Research Board (SERB), Government of India for providing financial support under the SERB Research Scientist scheme, Grant No. SB/SRS/2019-20/03/ES.

## Declarations

**Funding**

This work was funded by Science and Engineering Research Board (SERB), Government of India, Grant No. SB/SRS/2019-20/03/ES.

**Data availability statement**

The datasets generated during and/or analysed during the current study are available from the corresponding author on reasonable request.

**Conflicts of interest/Competing interests**

The author declares there are no conflicts of interests / competing interests.

**Code availability**

The calculations in this work were carried out using Quantum ESPRESSO package as implemented in Materials Square. (Available at: https://www.materialssquare.com/)
For visualization of cube files for ELF and total potential, VESTA software was used. (Available at: https://jp-minerals.org/vesta/en/download.html )